# Crystal growth and detector performance of large size high-purity Ge crystals

Guojian Wang[1*], Mark Amman[2], Hao Mei[1], Dongming Mei[1], Klaus Irmscher[3], Yutong Guan[1], Gang Yang[1]

1. Department of Physics, University of South Dakota, Vermillion, South Dakota, 57069, USA
2. Ernest Orlando Lawrence Berkeley National Laboratory, University of California, Berkeley, California, 94720, USA
3. Leibniz Institute for Crystal Growth, Max-Born-Straße 2, 12489 Berlin, Germany

* Corresponding author email: Guojian.Wang@usd.edu

**Abstract**

High-purity germanium crystals approximately 12 cm in diameter were grown in a hydrogen atmosphere using the Czochralski method. The dislocation density of the crystals was determined to be in the range of 2000 - 4200 $cm^{-2}$, which meets a requirement for use as a radiation detector. The axial and radial distributions of impurities in the crystals were measured and are discussed. A planar detector was also fabricated from one of the crystals and then evaluated for electrical and spectral performance. Measurements of gamma-ray spectra from Cs-137 and Am-241 sources demonstrate that the detector has excellent energy resolution.

## 1. Introduction

Recently, low background high-purity germanium (HP-Ge) crystals, with impurity levels of 0.5 - $2\times10^{10}cm^{-3}$, have attracted great attention owing to their application as detectors of rare event physics phenomena including the search for dark matter [1-3]. Several research projects, including Super-CDMS, EDELWEISS, GERDA, MAJORANA, CDEX [4-9], with large collaborations, have been designed to detect dark matter or neutrinoless double-beta decay utilizing large-scale HP-Ge detectors with ultra-low internal radioactive backgrounds. The sensitivity of germanium-based experiments is related to the size of the HP-Ge crystals. Large diameter (>3 inch) HP-Ge crystals have higher sensitivity and larger volume for detecting events, particularly from weakly interacting massive particles (WIMPs), a plausible dark matter candidate.

HP-Ge crystals grown on the earth surface environment without shielding from cosmic rays, can result in detectors that are contaminated by long-lived radio isotopes [10]. The long-lived isotopes can limit the detector's sensitivity. To minimize these unwanted long-lived isotopes, the best solution is to zone-refine germanium ingots, grow crystals and fabricate detectors in an underground environment where cosmic rays are largely shielded by the rock overburden. With reduced cosmic ray exposure, the HP-Ge detectors will be preferable for dark matter and neutrinoless double-beta decay experiments. The growth of HP-Ge crystals in an underground environment is therefore a critical step for underground experiments.

The growth of detector-grade HP-Ge crystals has been described by many authors [11-15]. Numerous steps are required to convert bulk germanium into the ultra-pure material used in detectors for rare event experiments. For detector fabrication, the HP-Ge crystal should be grown in a highly pure hydrogen atmosphere [11, 12] and the dislocation density should be in the range of $10^2$ - $10^4$ $cm^{-2}$ to avoid charge trapping from di-vacancy hydrogen ($V_2H$), and the net carrier concentration should be at a level of approximately $10^{10}$ $cm^{-3}$ [13-15]. Ge crystals can be grown up to 30 cm in diameter that are dislocation free [16]. These types of crystals are usually grown in an inert atmosphere with resistance heating; conditions more easily controlled when compared to crystal growth in a high-purity hydrogen atmosphere using induction heating. However, these large diameter, dislocation-free crystals typically have impurity levels and degraded charge transport, thereby making them inappropriate for use in high-energy-resolution radiation detection. Up to now, the largest commercially available detector-grade HP-Ge crystals are only ~10 cm in diameter. Our ultimate goal is to grow large diameter (~15cm) detector-grade HP-Ge crystals in an underground environment with the already mentioned advantages. However, much research and development must be performed on the surface in order to establish a detector-grade crystal growth process prior to moving the crystal growth facility underground. For example, issues related to the diameter enlargement during crystal growth, e.g. maintaining control over purity and dislocation density can more easily be solved in a surface laboratory than one underground. Once our crystal growth process is established and moved underground, the resultant large diameter HP-Ge crystals would be used for the detection of rare event physics within underground laboratories.

The growth of 3 - 9 cm diameter HP-Ge crystals has been reported previously by our lab [17, 18]. Here, we report on the large diameter (approximately 12 cm) HP-Ge crystal growth, as well as the impurity distributions and the detector performance obtained from samples cut from a large diameter crystal.

## 2. Experimental methods

The <100> oriented HP-Ge single crystals were grown by the Czochralski method in a hydrogen atmosphere [17, 18]. Before each crystal growth, the quartz crucible, the raw Ge ingot and the Ge seed were cleaned using the procedure described by Hansen [19].The Ge ingot was loaded into the quartz crucible with a diameter of 18.5 cm and a height of 11 cm. The quartz crucible was then placed into a highly pure graphite susceptor, which was used to convert the radio frequency energy into the heat needed to melt the Ge ingot. The dash necking process was used to reduce the dislocation density. At the completion of the crystal growth, all of the Ge melt was pulled out. If this was not done, the remaining melt would break the quartz crucible when the Ge melt solidified.

Figure 1 shows one of the crystals grown using the process described above. The crystal has a diameter of nearly 13 cm and a weight of 6.2 kg. There are four (111) facets on the shoulder of crystal, which were used to orient the crystal by a laser reflection method, developed in our lab [20].

Following the crystal growth, samples were cut from the oriented crystal using a diamond wire saw. These samples were then used for X-ray diffraction, dislocation, Hall effect, photo-thermal ionization spectroscopy (PTIS), and radiation detection measurements. For the X-ray diffraction measurements, two thin square samples (1.5x1.5x0.16 $cm^3$) were cut from the center of wafers taken from near the top and the bottom of the No.20 crystal. X-ray rocking curves, which provide a measure of crystalline quality, were measured on these samples using a Rigaku Ultima IV X-ray diffractometer.

Dislocation measurements were made on wafers cut from the top, middle, and bottom of crystal No.20. After lapping, polishing and etching, the dislocations on the wafers were counted using an Olympus BX40 microscope [18].

The net carrier concentration and material typeness of the samples were determined using Hall effect measurements. For these measurements, square samples with dimensions of 1.5x1.5x0.16 $cm^3$ were cut from wafers belonging to different axial positions of the crystals. These samples were polished, ultrasonically washed in a methanol, rinsed with deionized water, and dried with nitrogen gas. In-Ga eutectic ohmic contacts of a very small size compared to the sample dimensions were then scratched onto the four corners of the samples. Finally, the samples were measured at 77 K using the Hall effect with the Van der Pauw geometry. An Ecopia HMS-3000 with a 0.55 T permanent magnet was used to make the measurements.

In order to identify the electrical impurities present in the samples, photo-thermal ionization spectroscopy [13] was used. This was accomplished with a Fourier-transform infrared spectrometer (Bruker Vertex 80v) equipped with a liquid helium flow cryostat (Oxford CF 1204D).

The ultimate test of the quality of a HP-Ge crystal is to produce a detector from the material and evaluate it for electrical and spectral performance. To convert a sample into a detector for such testing, the following procedure was used. The sample was first cut to a shape whose cross-section is shown in Figure 2. The thin extension at the bottom of the sample was used for handling during the detector fabrication and, during detector operation, remained undepleted. Following the cutting, the sample was lapped and then chemically polished in a 4:1 nitric to hydrofluoric acid mixture. After the mechanical and chemical processing, electrical contacts were formed on the sample. Amorphous semiconductor based electrical contacts [21-26] were RF sputter deposited with the resultant structure schematically shown in Figure 2. The first of two sputter depositions consisted of coating the top contact face and the sides of the crystal with amorphous Ge (a-Ge). The second deposition coated the bottom with amorphous Si (a-Si). Both depositions used an Ar-$H_2$ (7% $H_2$) gas mixture for the sputter gas. With the sample completely coated with amorphous semiconductor, Al was thermally evaporated onto both the top and bottom surfaces in order to complete the detector.

Following fabrication, the detector was loaded into a vacuum cryostat and cooled to 79 K. Leakage current and detector capacitance were measured as a function of the voltage bias applied to the detector. Following this, the gamma-ray spectral performance of the detector was evaluated using Cs-137 and Co-60 sources. Signals from the detector were read out using a charge sensitive preamplifier operating at room temperature followed by commercial pulse processing electronics consisting of an analog pulse-shaping amplifier and a multichannel analyzer.

## 3. Results and discussion

### 3.1 Structural quality of the crystals

X-ray rocking curves obtained from measurements made on two samples cut from HP-Ge crystals No.20 as shown in Figure 3. The sharp and distinct diffraction peaks at $32.9997^{o}$ with FWHM = 0.038 - $0.040^{0}$ confirm the high crystalline quality of the HP-Ge samples.

For dislocation measurements, in order to have an accurate count of the dislocations, 41 measurement zones (area 300 x 400 $\mu m^2$) on each wafer were counted. The average dislocation densities for the top, middle, and bottom samples obtained from crystal No.20 were 2000, 3400, and 4200 $cm^{-2}$, respectively (as shown in Fig. 4). These values fall within the range required for use of the HP-Ge as a radiation detector.

### 3.2 Radial distribution of the impurities

In HP-Ge, the dominant acceptors are boron, aluminum and gallium, while the dominant donor is phosphorus [27]. Boron and aluminum will react with the silica crucible to form oxidation complexes, making the effective segregation coefficient of aluminum approach one and boron to disappear faster than its segregation coefficient would predict. In a hydrogen atmosphere, no oxidation complexes exist in Ge crystals for gallium and phosphorus. Therefore, gallium and phosphorus can segregate normally.

However, the quartz crucible, insulation materials and a poor vacuum can bring phosphorus contamination into the Ge melt and the phosphorus distribution in the crystal may deviate from that predicted by the segregation coefficient as reported in [12, 27, and 28].

For use as a radiation detector, the net carrier concentration of HP-Ge should be less than about $2x10^{10}$ $cm^{-3}$, so that depletion lengths one or more centimeters in size can be achieved at reasonable applied voltages (< 5 kV). The net carrier concentration ($N_{net}$) can be defined as below [29]:

$$N_{net}= N_A-N_D= 1/(\rho e \mu_{(n\ or\ p)})$$

where $N_A$ is the concentration of acceptors and $N_D$ is the concentration of donors, $\rho$ is the resistivity, e is the electron charge, $1.6x10^{-19}$ C, and $\mu$ is the drift mobility associated with the carrier type of the sample, either n or p. The net carrier concentration can be measured using the Hall effect. In order to get reasonable net carrier concentrations, the measured Hall mobility should be higher than 25,000 $cm^2V^{-1}s^{-1}$ at 77 K [30]. For our measurements, the Hall mobility was in the range of 30,000 - 45,000 $cm^2V^{-1}s^{-1}$.

Figure 5 (right) shows the radial distribution of carrier concentrations for three wafers cut from crystal No. 20. Over the range of g = 0.06 - 0.2, where g is the fraction of the melt that has crystallized, the net carrier is p type and the core of the crystal shows lower carrier concentration than the edges. In order to verify the main impurities leading to the distributions of Figure 5, samples from the center and edge of the wafer at g = 0.1 were prepared for PTIS measurements. The PTIS results are shown in Figure 6. For the center sample, the main impurity is aluminum. For the edge sample, the main impurities are aluminum and gallium. Additionally, small amounts of boron and phosphorus can be detected. In both samples compensating donors of unknown origin are observed having absorption lines in the range from 90 to 300 $cm^{-1}$. These features will be the subject of further investigations.

Based on the observed shape of the interface between the crystal and the melt during crystal growth, the segregation coefficients of the impurities, and the PTIS measurements, the radial distributions of Figure 5 can be understood. The melt and crystal interface is convex to the melt, and, at this interface, the impurities are expected to be uniformly distributed. As a result of the interface shape, the center of a sample wafer cut from a crystal would have solidified before the edge. For the acceptor impurities of boron, aluminum, and gallium, the effective segregation coefficients are 9.5, 1 and 0.1, respectively [27]. Therefore, aluminum will distribute uniformly in the germanium crystal, boron tends to accumulate to the top part of crystal and gallium will be concentrated at the bottom part of crystal. Based on this and the PTIS measurements previously shown, we conclude that the non-uniform distribution of holes shown in Figure 5 (right) is mainly caused by the distribution of gallium.

Figure 7 shows the distribution of the net carrier concentration along the radial direction for crystals No.20 and 34. An approximate 8 cm core of each crystal exhibits a uniform impurity concentration along the radial direction. This means the crystal-melt interface is almost flat for 8 cm. A uniform distribution of impurities along the radial direction is desirable for radiation detection applications.

**3.3 Axial distribution of the impurities**

Figure 8 presents the axial net carrier concentration distribution of central samples from the head to the tail of crystals No. 20 and 34. From the head to 30% weight, the crystals are p-type with the carrier concentration in the range of 2 - $8x10^{10}$ $cm^{-3}$. A sample cut from this region of crystal No.20 was made into a planar detector and will be discussed in the following section. Beyond 30% of crystal, the crystal changed from p-type to n-type and the dominant impurity was phosphorus [12, 13 and 27].

**3. 4 Detector performance**

In order to assess the quality of our HP-Ge crystals, a sample was cut from the No. 20 crystal (as shown in Fig.2 (b)), converted into a detector, and evaluated for electrical and spectral performance as described previously. The resultant detector had an active area of 11.3 mm by 11.6 mm and a thickness of 7.7 mm. The detector exhibited a low leakage current of less than 1 pA at 3000 V. The capacitance data obtained with the detector is plotted in Figure 9. It is evident from the plot that full depletion of the detector is not achieved since the capacitance never becomes constant with increasing detector voltage. A linear fit to the data shown in Figure 9 allows the net ionized impurity concentration of the sample to be estimated. The slope of the fit is equal to $2/A^2\varepsilon eN$, where A is the detector area, ε is the dielectric constant of Ge, e is the magnitude of the electron charge, and N is the net ionized impurity concentration. Using this equation and the extracted slope, we obtain $1x10^{11}/cm^3$ for the impurity concentration. It should be noted that the net carrier concentration determined through the Hall effect was 2 – 6 x $10^{10}/cm^3$ for this sample. The origin of this difference between the Hall effect and C-V measurements is not clear, and more crystals will be compared in future to address this issue.

Spectra from Cs-137 and Am-241 sources were acquired with the detector at a detector bias voltage of 3000 V and are shown in Figures 10 and 11. The net energy resolutions of 1.24 keV FWHM at 662 keV and 1.51 keV FWHM at 59.5 keV are achieved with this sample. For comparison, the FWHM peak broadening expected from charge generation statistics is given by $2.35(F\delta E)^{1/2}$, where F is the Fano factor, $\delta$=2.96 eV is the average energy to create an electron-hole pair, and E is the gamma-ray energy [31]. At the 662 keV gamma-ray energy and assuming a Fano factor of 0.08, a statistical broadening of 0.93 keV is predicted. Though the net resolution obtained with our detector at 662 keV is a bit above this value, the resolution is comparable to that obtained with commercially-grown HP-Ge and indicates that our HP-Ge has excellent charge transport. The primary remaining challenge is to lower the net impurity concentration so that large depletion widths can be realized.

**4. Conclusions**

HP-Ge crystals with diameters of approximately 12 cm were grown by the Czochralski method. The crystals were of high quality with dislocation densities in the range of 2000 - 4200 $cm^{-2}$. A non-uniform distribution of p-type carriers along the radial direction was observed and determined to be caused by gallium. For two crystals, the low point of the net carrier concentration measured along the length of the crystals was in the range of 2 - $8x10^{10}$ $cm^{-3}$. A detector was fabricated from a sample cut from one of these two crystals, and the spectral performance of the detector was evaluated using Cs-137 and Am-241 sources. The detector measurements demonstrated that excellent energy resolution can be achieved with our HP-Ge crystals.

**Acknowledgement**

We would like to thank Angela Chiller for a carefully reading of this manuscript and Mike Pietsch for assistance in PTIS measurement. This work is supported by DOE grant DE-FG02-10ER46709 and the state of South Dakota.

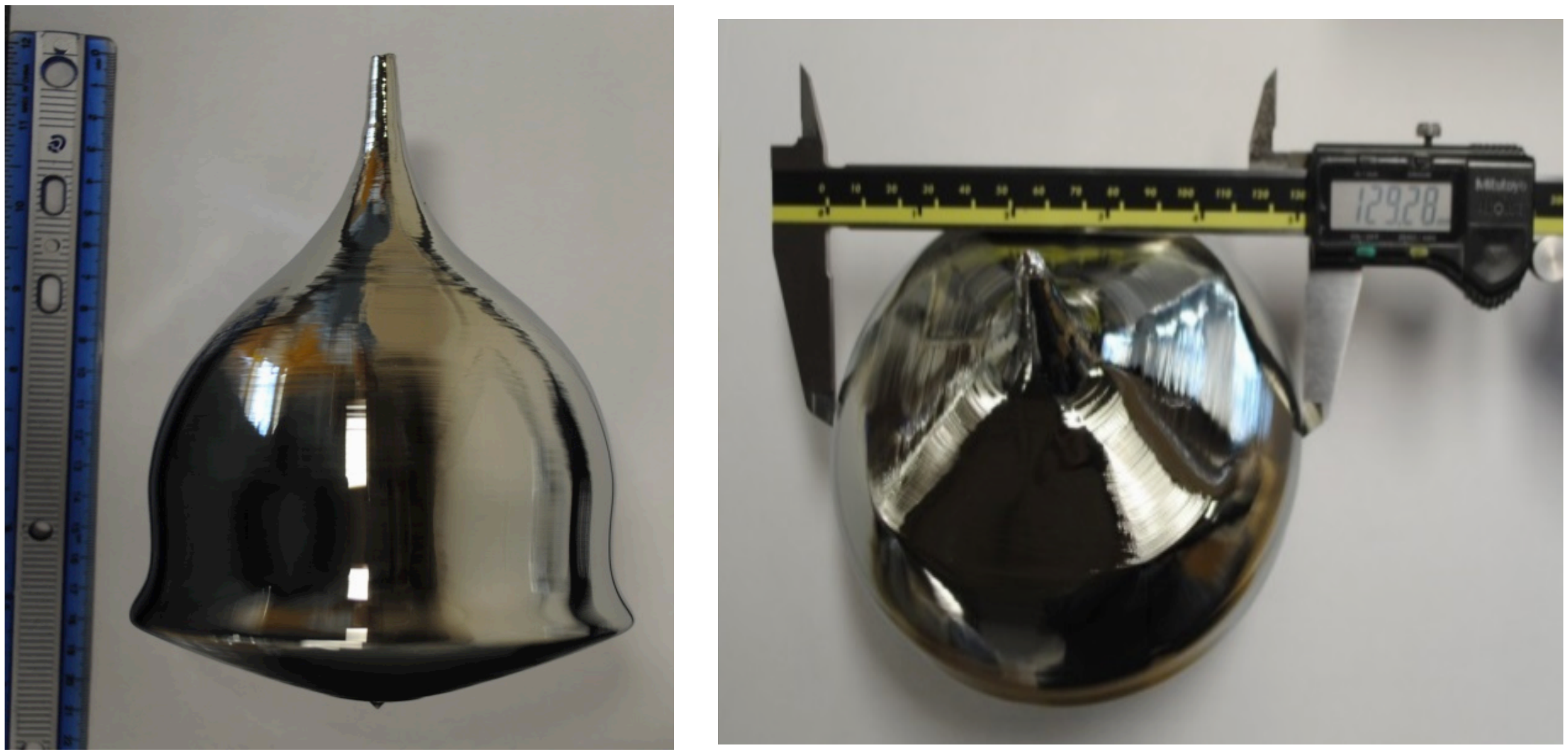

Fig.1 HP-Ge crystal (No.20)

(a)

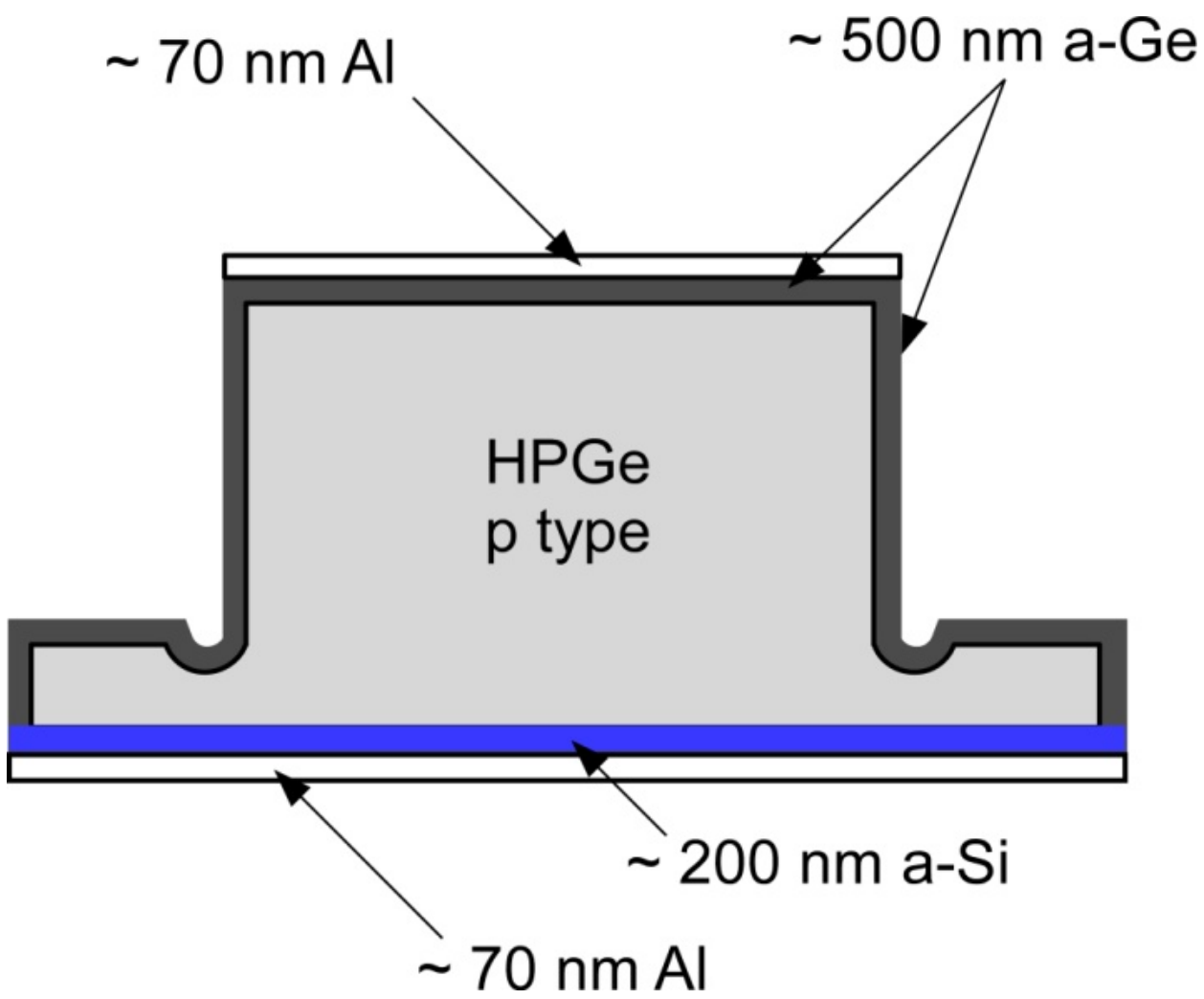

(b)

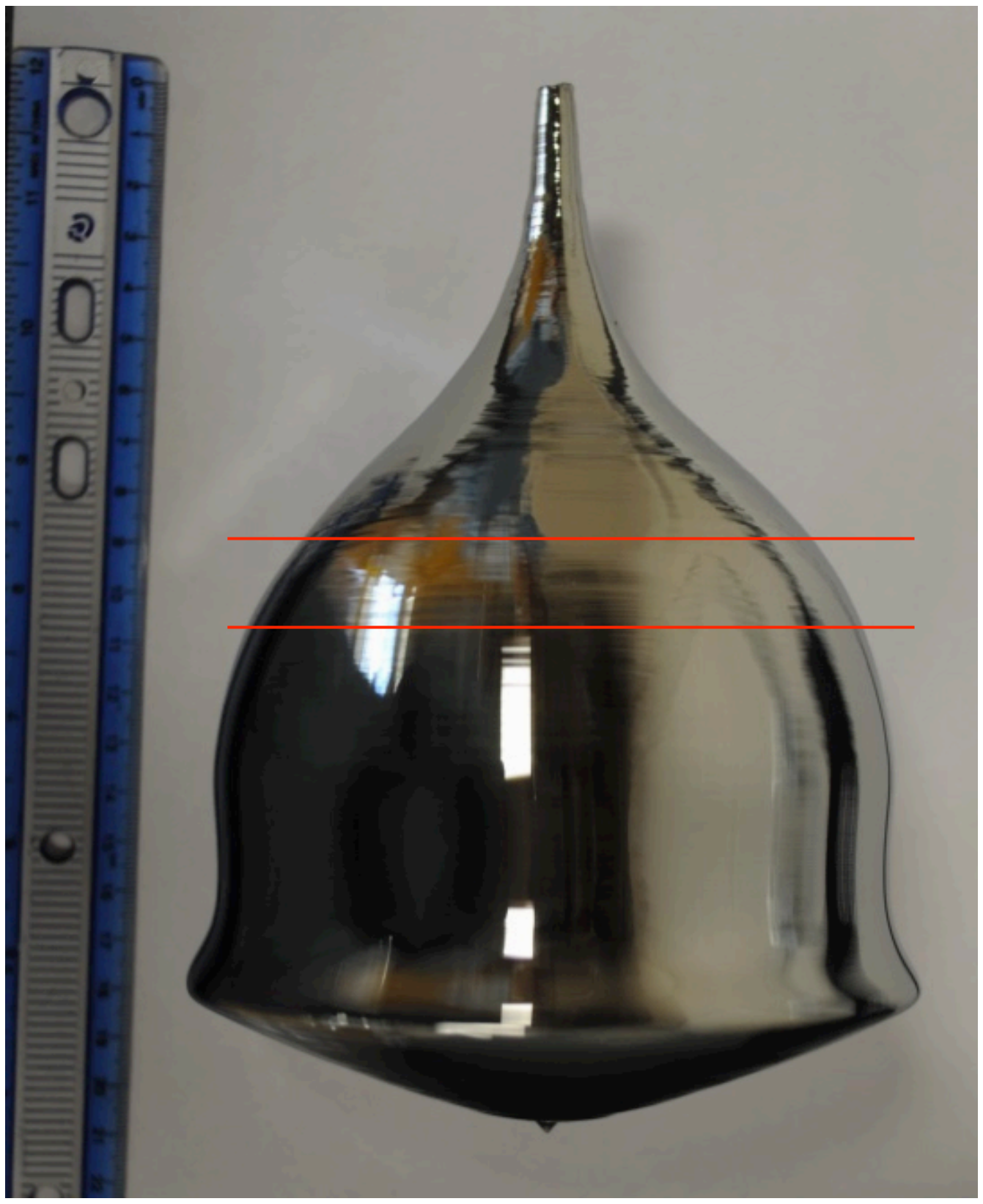

Fig. 2 (a) Schematic cross-sectional drawing showing the geometry of the HP-Ge detector.

(b) Photograph of the crystal from which the detector was cut. The red lines designate the slice that was used for the detector.

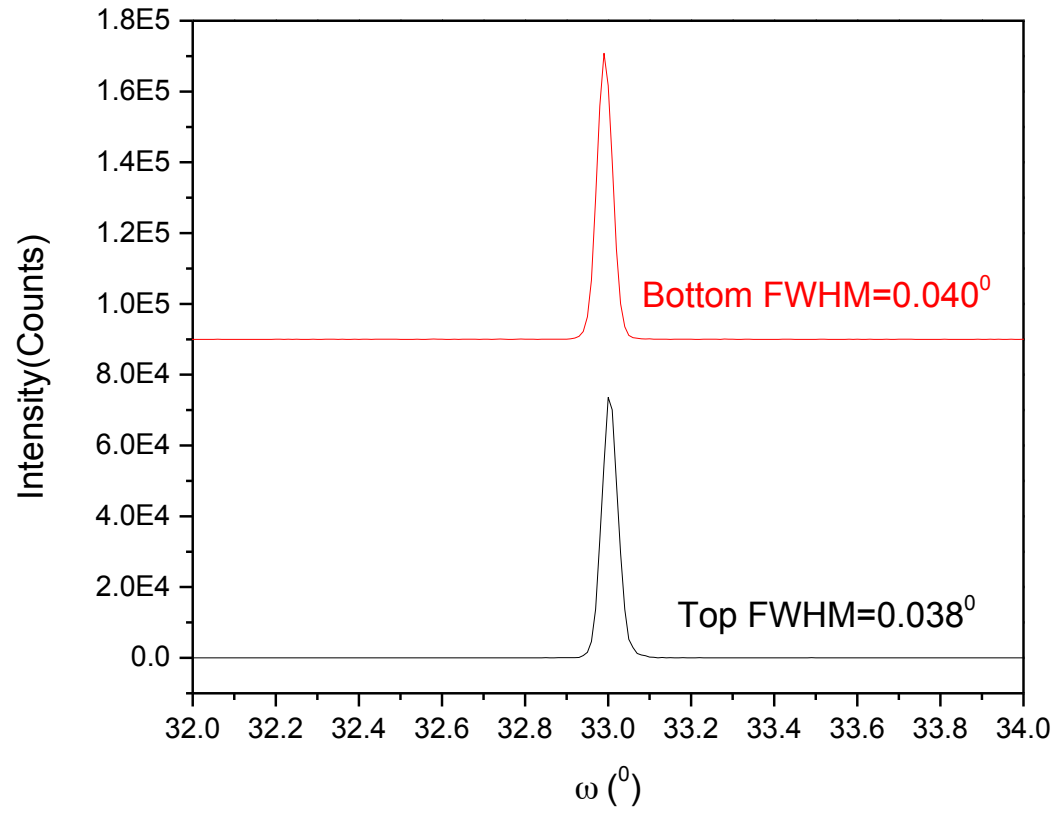

Fig. 3 X-ray rocking curves of the (100) plane obtained from samples cut from the HP-Ge crystal No.20.

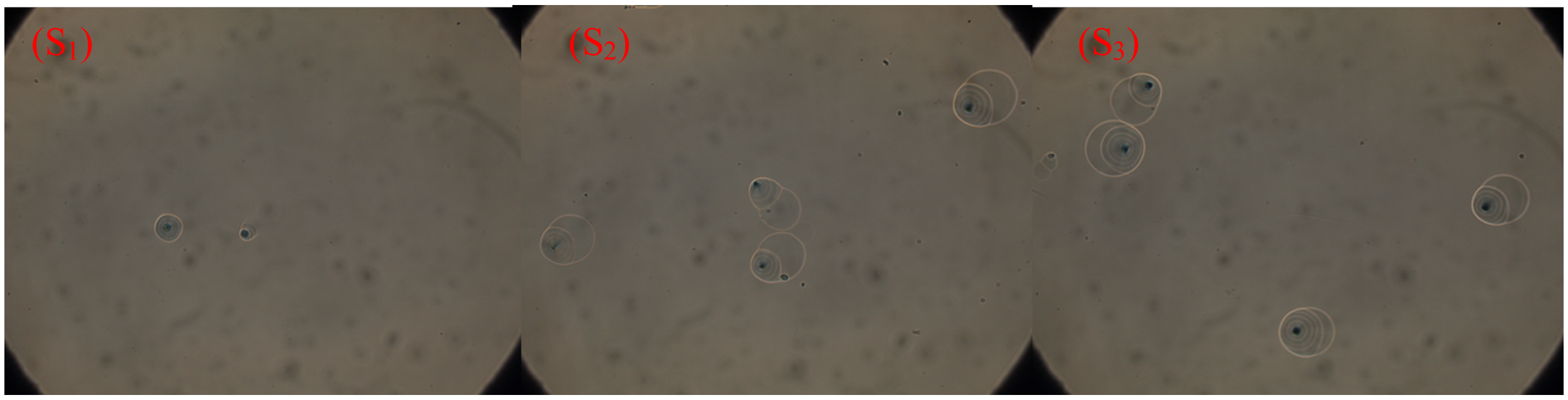

Fig. 4. Dislocations in the top ($S_1$), middle ($S_2$) and bottom ($S_3$) portions of the HP-Ge crystal No.20. The area of the images is 300 x 400 μm$^2$. The dislocation densities are 2000, 3400 and 4200 cm$^{-2}$ for $S_1$, $S_2$, and $S_3$, respectively.

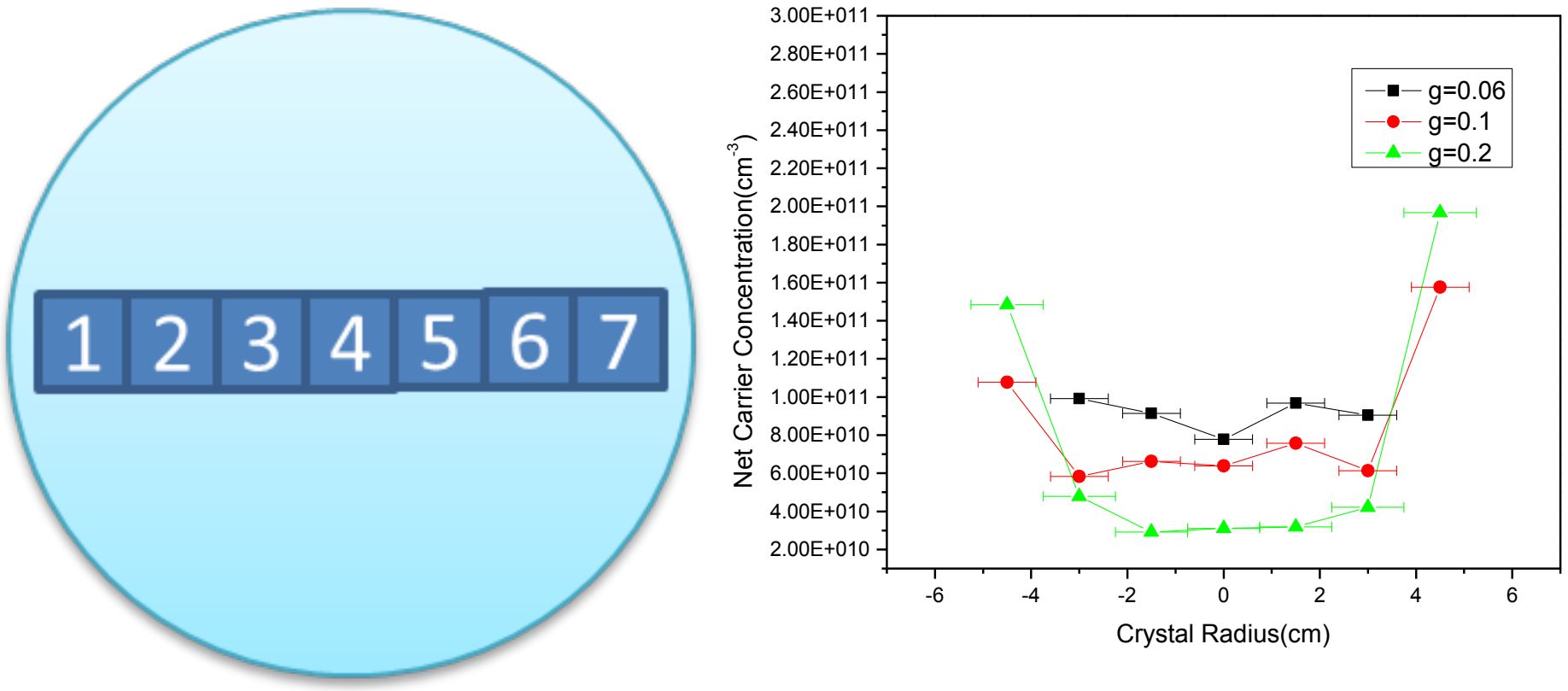

Fig. 5 Schematic diagram showing the location of the samples in the wafers for the Hall effect measurements (left) and radial distribution of the hole concentration measured at 77 K for three wafers cut from different axial positions (denoted by g, which is the fraction of the melt that has crystallized) of crystal No.20 (right).

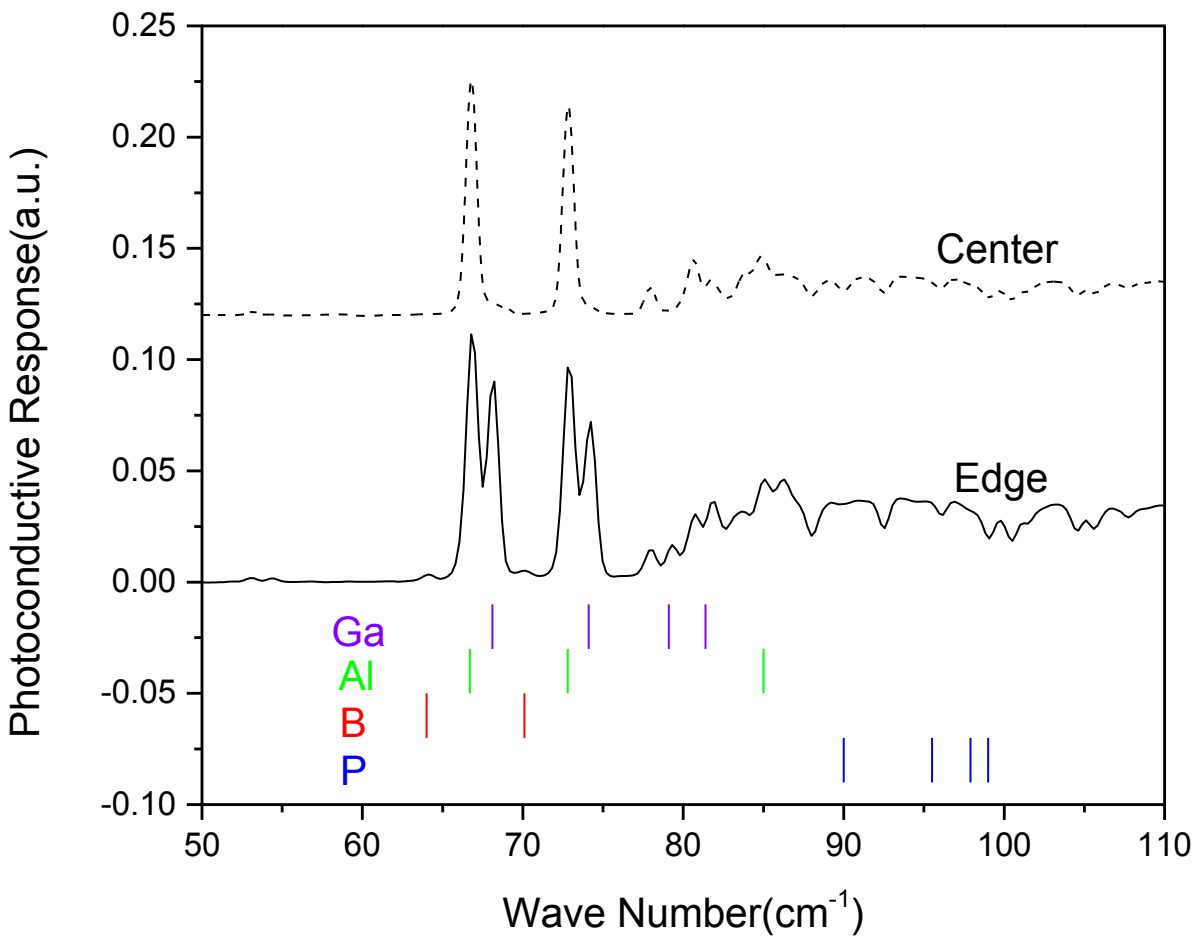

Fig. 6 Photo-thermal ionization spectra at 7 K from the center and the edge of a wafer (g=0.1) for crystal No. 20.

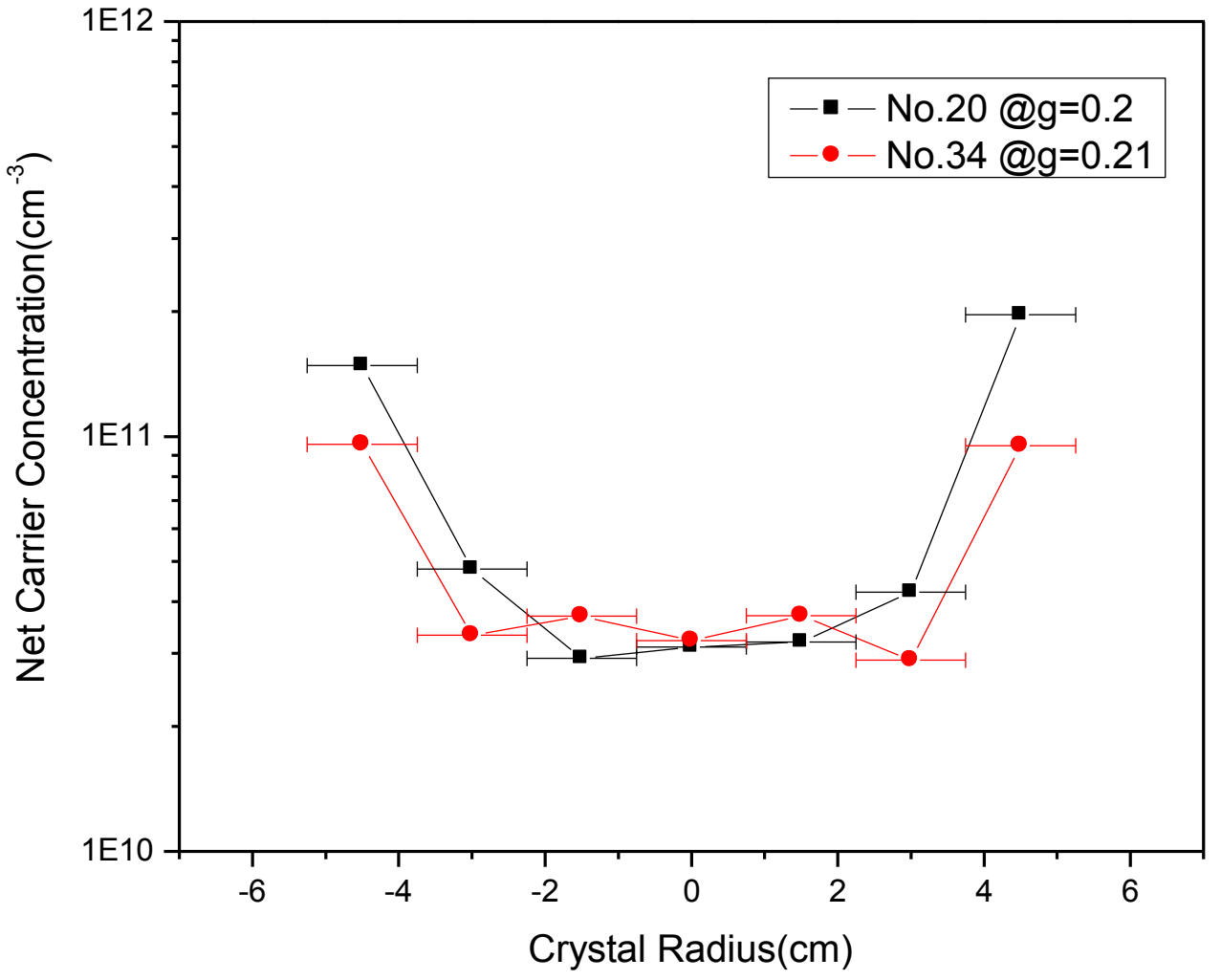

Fig. 7 Measured radial distribution of the net carrier concentration in crystals No.20 and 34.

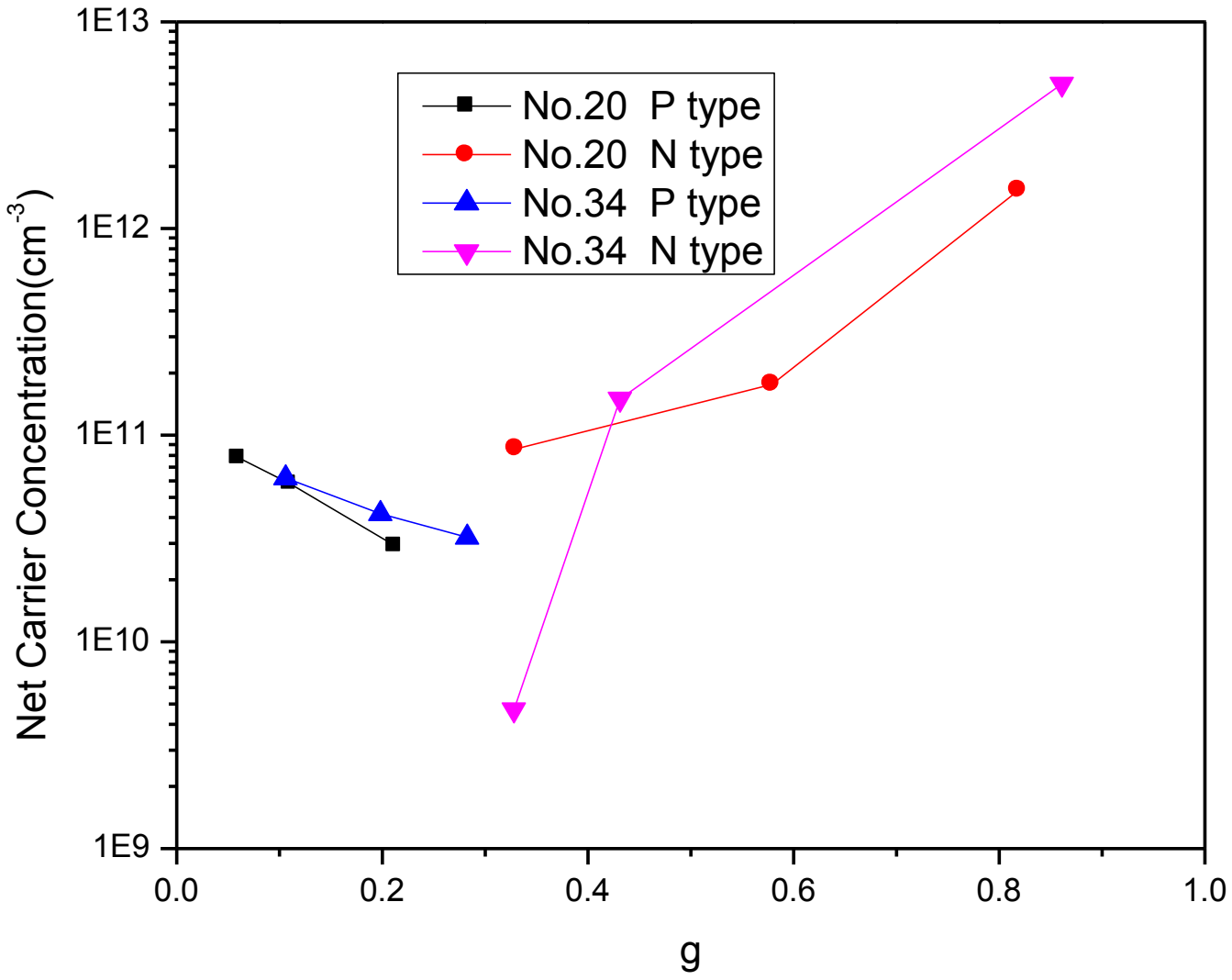

Fig. 8 Measured axial distribution of the net carrier concentration in crystals No. 20 and 34, where g is the fraction of the melt that has crystallized.

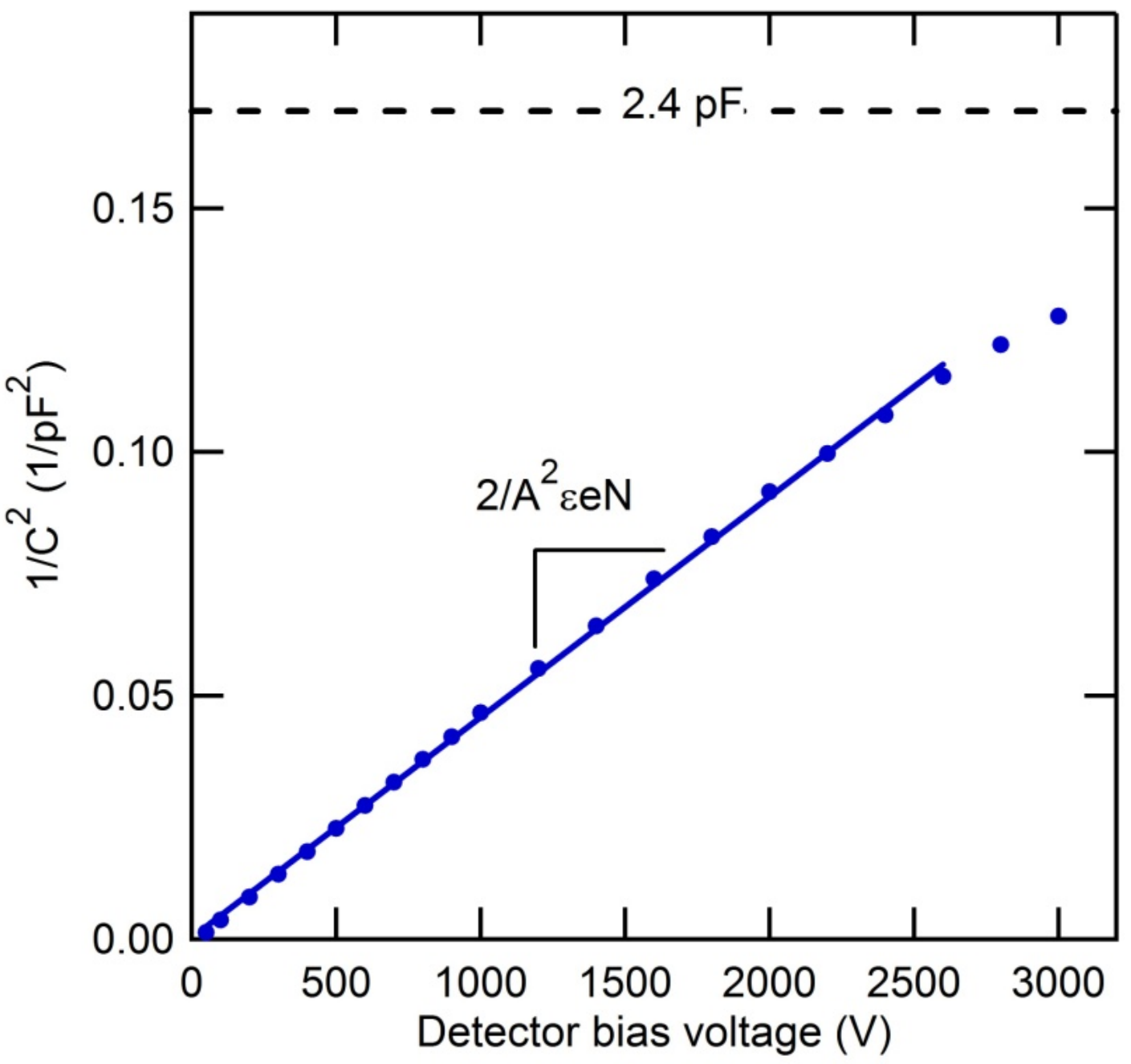

Fig. 9 Inverse of the measured detector capacitance squared plotted as a function of bias voltage for a HP-Ge detector of the configuration shown in Figure 2. The circles are the measured data and the solid line is a fit to the data. The dashed line corresponds to the expected fully depleted detector capacitance of 2.4 pF.

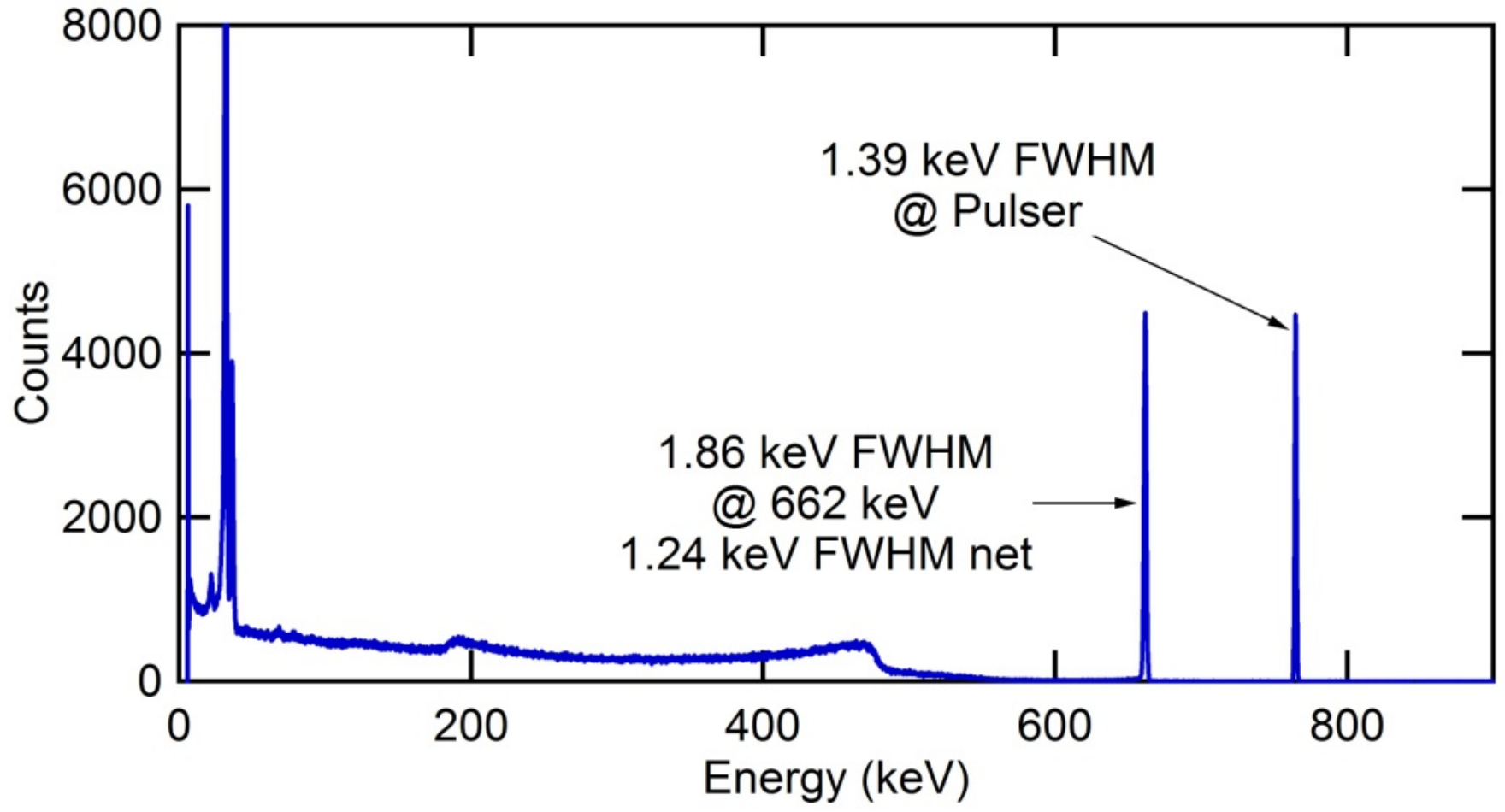

Fig. 10 Energy spectrum of Cs-137 measured with a HP-Ge detector of the configuration shown in Figure 2. The detector was operated with a voltage bias of 3000 V, and a pulse peaking time of 8 μs was used for the measurement.

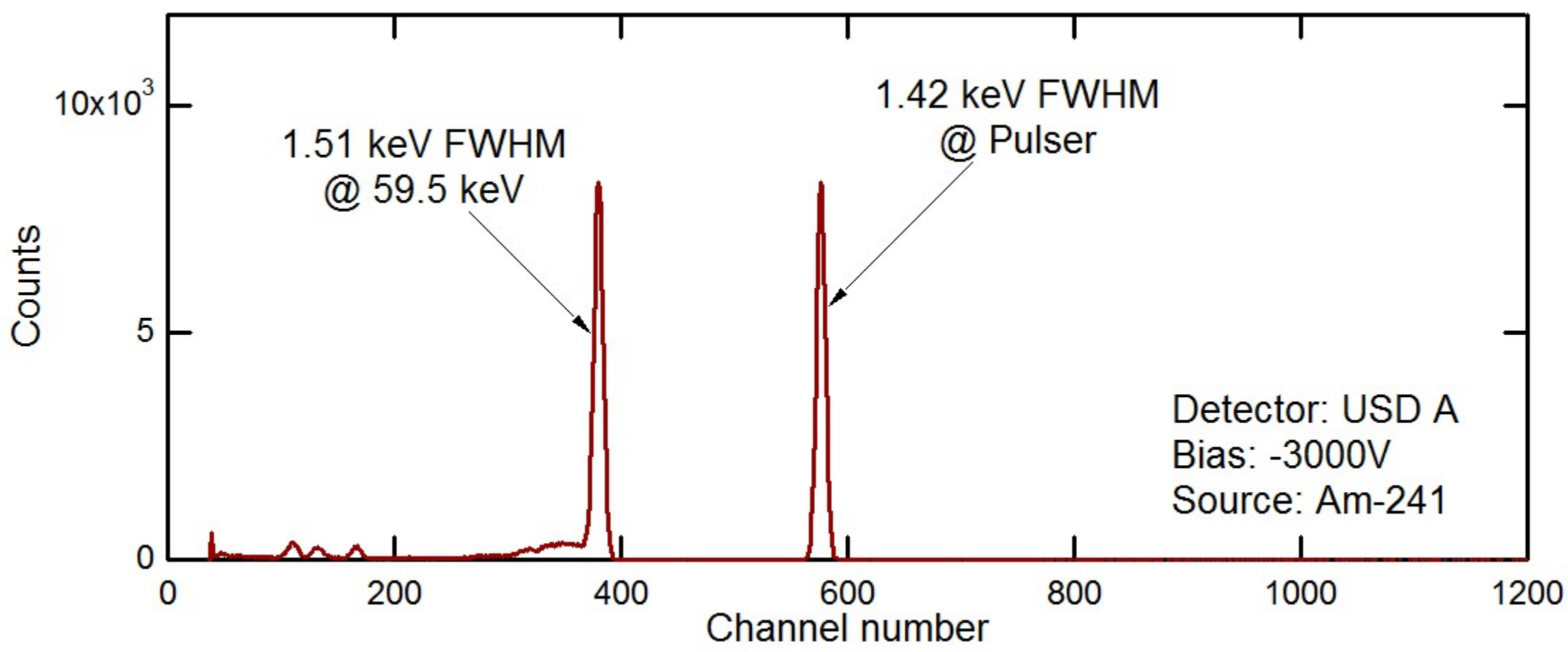

Fig. 11 Energy spectrum of Am-241 measured with a HP-Ge detector of the configuration shown in Figure 2. The detector was operated with a voltage bias of 3000 V, and a pulse peaking time of 8 μs was used for the measurement.